\documentclass[
amssymb,amsmath,aps,prl
,reprint 
]{revtex4-1}

\usepackage{docs}%
\usepackage{bm}%
\usepackage{graphicx}
\usepackage{subfigure}
\usepackage{booktabs}
\expandafter\ifx\csname package@font\endcsname\relax\else
 \expandafter\expandafter
 \expandafter\usepackage
 \expandafter\expandafter
 \expandafter{\csname package@font\endcsname}%
\fi
\draft 

\begin{document}

\title{\boldmath Observation and calculation of the quasi-bound rovibrational levels of the electronic ground state of H$_2^+$}

\author{Maximilian Beyer}
\affiliation{Laboratorium f\"ur Physikalische Chemie, ETH Z\"urich, 8093 Z\"urich, Switzerland}
\author{Fr\'ed\'eric Merkt}
\affiliation{Laboratorium f\"ur Physikalische Chemie, ETH Z\"urich, 8093 Z\"urich, Switzerland}

\date{\today}

\begin{abstract}
Although the existence of quasi-bound rotational levels of the $X^+ \ ^2\Sigma_g^+$ ground state of H$_2^+$ has been predicted a long time ago, these states have never been observed. Calculated positions and widths of quasi-bound rotational levels located close to the top of the centrifugal barriers have not been reported either. Given the role that such states play in the recombination of H(1s) and H$^+$ to form H$_2^+$, this lack of data may be regarded as one of the largest unknown aspects of this otherwise accurately known fundamental molecular cation. We present measurements of the positions and widths of the lowest-lying quasi-bound rotational levels  of H$_2^+$ and compare the experimental results with the positions and widths we calculate using a potential model for the $X^+$ state of H$_2^+$ which includes adiabatic, nonadiabatic, relativistic and radiative corrections to the Born-Oppenheimer approximation.
\end{abstract}


\maketitle

The theoretical treatment of H$_2^+$ has played and is still playing an important role in the development of quantum chemistry. H$_2^+$ possesses one electron and its energy-level structure can be calculated with extraordinary accuracy. Highly accurate ab-initio calculations started with the nonadiabatic theory of Kolos and Wolniewicz  \cite{kolos63a} and its applications to H$_2$ \cite{kolos93a,wolniewicz93a,kolos94a,wolniewicz95a,piszczatowski09a} and H$_2^+$ \cite{bishop73a,bishop77a,bishop78a, wolniewicz86a}. To compute nonadiabatic effects in H$_2^+$, approaches based on a transformed Hamiltonian \cite{bunker77a,moss93a} and coordinate-dependent vibrational and rotational masses \cite{herman66a,schwenke01a,kutzelnigg07a, jaquet08a,diniz12a}  are particularly successful. Promising alternative methods of computing the energy-level structure of H$_2^+$ and H$_2$ not relying on the Born-Oppenheimer approximation have also been developed  \cite{taylor99a,korobov00a,hilico00a,matyus12a,stanke13a}. High-order perturbative calculations of relativistic and radiative corrections have been reported both for H$_2^+$ \cite{bukowski92a,korobov06a,korobov08a,korobov14a} and H$_2$ \cite{piszczatowski09a,komasa11a}. 

H$_2^+$, HD$^+$, H$_2$, and HD are among the first molecules to have been formed in the universe and are therefore also of central importance in astrophysics. Through reactions with H$_2$, the most abundant molecule in the interstellar medium, H$_2^+$ is converted into H$_3^+$, so that H$_2^+$ has not been detected in astrophysical spectra so far \cite{ehrenfreund00a,hirata06a}, despite extensive searches by radioastronomy.

 481 and 4 rovibrational levels are believed to exist in the ground ($X^+\ ^2\Sigma_g^+$) and first excited ($A^+\ ^2\Sigma_u^+$) electronic states of H$_2^+$, respectively \cite{peek69a,moss93a,carbonell03a,carbonell04a}, but only a fraction of these have been observed experimentally, using methods as diverse as microwave electronic \cite{carrington89a, carrington89b,carrington93a} and pure rotational \cite{critchley01a} spectroscopy,
radio-frequency spectroscopy of magnetic transitions between fine- and hyperfine-structure components \cite{jefferts68a,jefferts69a}, photoelectron spectroscopy \cite{asbrink70a,merkt92a,chang07a}, and Rydberg-state spectroscopy combined with Rydberg-series extrapolation \cite{herzberg72a,arcuni90a,osterwalder04a,liu09a,haase15a}. The main reason for the incomplete experimental data set on the level structure of H$_2^+$ is the absence of allowed electric-dipole rotational and vibrational transitions.

Of the 481 rovibrational levels of the $X^+$ state of H$_2^+$, 58 are known to be quasi-bound tunneling (shape) resonances located above the H(1s) + H$^+$ dissociation limit, but below the maxima of the relevant centrifugal barriers \cite{moss93a}. Whereas 26 of these resonances are extremely narrow and can be calculated as accurately as bound levels, Moss lists 13 quasi-bound levels for which an accuracy of $10^{-4}$ cm$^{-1}$ could not be reached. 19 quasi-bound levels of the $X^+$ state are even located so close to the top of the respective centrifugal potential barriers that they could not be calculated so far  \cite{moss93a,jaquet08a}. These levels have not been observed experimentally either. This lack of knowledge is astonishing because shape resonances of H$_2^+$ are not only intrinsically interesting but also because they represent a channel for the formation of H$_2^+$ in H$^+$ + H(1s) collisions by radiative or three-body recombination.

We report the observation of the lowest-$N^+$ shape resonances of H$_2^+$, the $(v^+=18,N^+=4)$ resonance of para H$_2^+$ and the $(17,7)$ resonance of ortho H$_2^+$ and present calculations of their positions and widths using Born-Oppenheimer potential-energy functions of the $X^+$ state \cite{wind65a,peek65a,kolos69a,bishop73a}, and adiabatic \cite{kolos69a,bishop73a}, nonadiabatic \cite{jaquet08a},  relativistic and radiative corrections \cite{howells90a, moss93a, moss03a}.

\begin{figure}
  \includegraphics[width=\linewidth]{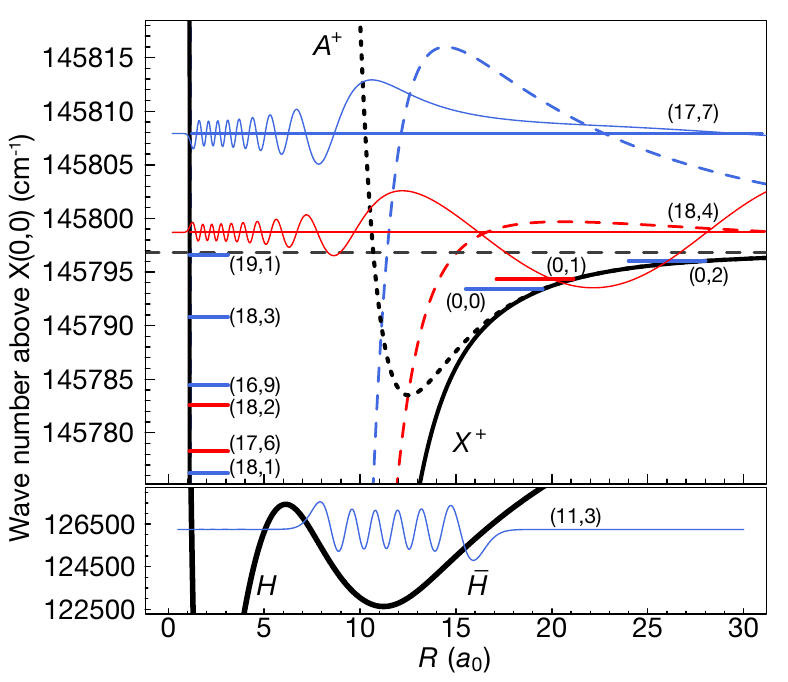}
 \caption{Potential-energy functions of the $H\bar{H}$ state of H$_2$~\cite{wolniewicz98b} (lower panel) and the $X^+$ (solid, dashed) and $A^+$ (dotted) states of H$_2^+$ \cite{bishop73a, peek65a}. Selected vibrational wave functions and energy levels of para (red) and ortho (blue) hydrogen are displayed.\label{fig1}}
\end{figure}

The quasi-bound levels of the $X^+$ state of H$_2^+$ were studied by pulsed-field-ionization zero-kinetic-energy (PFI-ZEKE) photoelectron spectroscopy \cite{muellerdethlefs98a} using an electric-field pulse sequence designed for high spectral resolution \cite{hollenstein01a}. The spectra were obtained by monitoring the electrons produced by field ionization of very high Rydberg states (principal quantum number $n$ beyond 100) located below the ionization thresholds of H$_2$ as a function of the wave number of a tunable laser.
To access the bound and quasi-bound rotational levels of the highest vibrational states ($v^+=16-19$) of the $X^+$ state of H$_2^+$ from the $X~^1\Sigma_g^+$ ground state of H$_2$, a three-photon excitation sequence 
\begin{align}
{\rm H}_2^+  &  \xleftarrow[+\rm{PFI}]{\rm{VIS2}}  \bar{H}(11,2\text{-}3) \xleftarrow{\rm{VIS1}} B(19, 1\text{-}2) \xleftarrow{\rm{VUV}} X(0, 0\text{-}1).  \label{eq1}
\end{align}  
was used via the  $B\ ^1\Sigma_u^+~(19,1$ or 2) and the $\bar{H}~^1\Sigma_g^+~(11, 2$ or 3) intermediate levels. Selecting vibrational levels of the outer ($\bar{H}$) well of the $H\bar{H}$ state is ideal for accessing long-range states of molecular hydrogen, as demonstrated by Reinhold {\it et al.}~\cite{reinhold97a,reinhold99a}, who also reported the absolute term values of many rovibrational levels of the $\bar{H}$ state. 
Fig.~\ref{fig1} depicts the potential-energy functions of the $H\bar{H}$ state \cite{wolniewicz98b} of H$_2$ (lower panel) and the $X^+$ ($N^+=0,4~\text{and}~7$) and $A^+$ states of H$_2^+$ \cite{bishop73a,peek65a} (upper panel) and selected vibrational wave functions. The figure indicates that the $v=11$ vibrational level of the $\bar{H}$ can be used to access the ionization continua associated with the highest vibrational levels of the $X^+$ and the few bound levels of the $A^+$ state.

The vacuum-ultraviolet (VUV) radiation around 105680~cm$^{-1}$ needed in the first step of the excitation sequence~(\ref{eq1}) was generated by four-wave mixing in a pulsed beam of Kr gas using two Nd:YAG-pumped pulsed dye lasers (pulse duration 5~ns), as described in Ref.~\cite{merkt98a}. A third pulsed dye laser was used to access the $\bar{H}~(v=11)$ levels from the selected levels of the $B$ state. A fourth tunable pulsed dye laser, delayed by approximately 10~ns with respect to the other two laser pulses, was used to access the region near the dissociative ionization (DI) threshold of H$_2$. 

\begin{figure*}
    \subfigure[\label{fig2a}]{\includegraphics[width=0.48\linewidth]{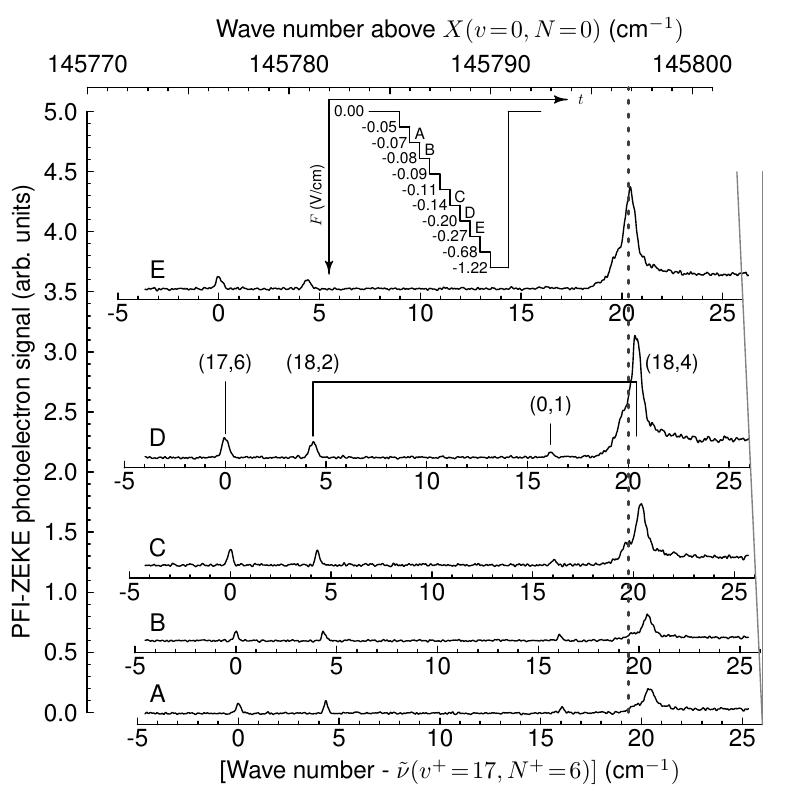}}
    \subfigure[\label{fig2b}]{\includegraphics[width=0.48\linewidth]{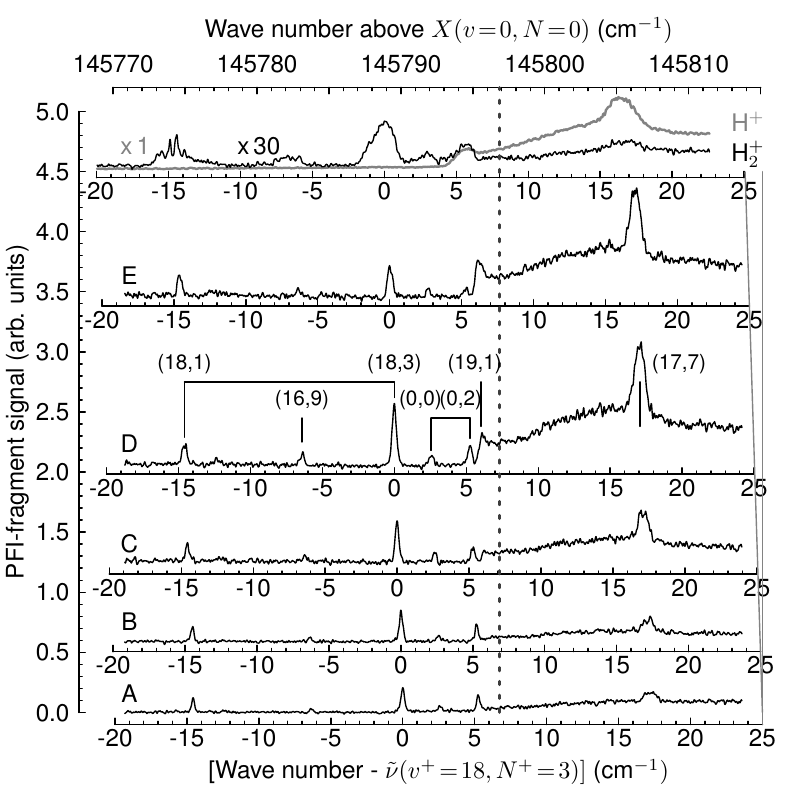}}

 \caption{PFI-ZEKE photoelectron spectra of para (a) and ortho (b) H$_2$ in the vicinity of the DI threshold, which is indicated by a grey vertical dashed line. The spectra were recorded with the electric field steps labeled A-E in the pulse sequence depicted in the inset in (a). MATI spectra for ortho H$_2$ are displayed in the upper trace in (b). \label{fig2}}
\end{figure*}

All three laser beams intersected a pulsed skimmed supersonic beam of neat H$_2$ at right angles on the axis of a PFI-ZEKE photoelectron spectrometer \cite{merkt98a}. The wave number of the fourth dye laser was calibrated at an accuracy of 600~MHz (3$\sigma$) using a wave meter. The resolution of the photoelectron spectra was determined by the bandwidth of about 1 GHz of the fourth dye laser and the selectivity of the PFI process. The instrumental line-shape functions adequate to describe the spectra recorded with the successive pulses of the PFI sequence (see inset of Fig.~\ref{fig2a}) are Gaussian functions with a full width at half maximum (FWHM) of 0.2~cm$^{-1}$ (pulses 2 to 5 in the pulse sequence), 0.25~cm$^{-1}$ (pulse 6), and 0.35~cm$^{-1}$ (pulses 7 and 8). The line widths of the spectra recorded with pulses 9 and 10 were too large, and the signal recorded with the first pulse was too weak, to be included in the analysis. Mass-analyzed threshold ionization (MATI) spectra~\cite{zhu91a} were recorded with a pulse sequence consisting of a discrimination pulse of $-70$~mV/cm followed by an extraction pulse of 800~mV/cm and monitoring H$^+$ and H$_2^+$ ions. 

Figures~\ref{fig2a} and~\ref{fig2b} display the PFI-ZEKE photoelectron spectra of para and ortho H$_2$ in the vicinity of the H$^+$ + H(1s) + e$^-$ DI threshold recorded from the $\bar{H}~^1\Sigma_g^+~(11, 2 \ {\rm and} \ 3)$ intermediate levels, respectively. Ten spectra were recorded simultaneously by monitoring the electrons produced by the ten electric-field steps of the pulse sequence (see inset of Fig.~\ref{fig2a}) but only five, corresponding to the steps labeled A-E, are shown for clarity. The upper horizontal scale indicates the wave number above the H$_2(v=0,N=0)$ ground state, which was determined from the known term value of the selected $\bar{H}~^1\Sigma_g^+~(11, N=2 \ {\rm or} \ 3)$ level \cite{reinhold99a} and the wave number $\tilde\nu_{\rm{VIS2}}$ (see Eq.~(\ref{eq1})). The scale given below each spectrum in Figs.~\ref{fig2a} and~\ref{fig2b} gives the wave number relative to the positions of the $X^+(17,6)$ and the $X^+(18,3)$ states, respectively. The position of the DI threshold, 145796.8413(4)~cm$^{-1}$ \cite{liu09a,mohr14a} is marked by a grey dashed vertical line and coincides with the onset of a continuum in the spectra. Because the spectra display the yield of electrons produced by delayed PFI, the electron signal measured in the continuum must stem from the field ionization of high-$n$ Rydberg states of H, a conclusion that was confirmed by the MATI spectra, displayed in the upper part of Fig.~\ref{fig2b}. 

The spectra of para H$_2$ (Fig.~\ref{fig2a}) consist each of three sharp lines located below the DI threshold, which can be unambiguously attributed to the (17,6) and (18,2) levels of the $X^+\  ^2 \Sigma_g^+$ ground state and the (0,1) level of the $A^+\  ^2\Sigma_u^+$ first excited state. The spectra also reveal a broader line above the DI threshold. Modeling the line shape by taking into account the experimental line-shape function and after subtraction of the contribution of the  DI continuum indicates a Lorentzian line-shape function with a FWHM of 0.21(7)~cm$^{-1}$, which suggests that this level is a quasi-bound level of H$_2^+$. Based on the calculations presented below, we assign this line to a transition to the quasi-bound (18,4) level of H$_2^+$. 

The spectra of ortho H$_2$ (Fig.~\ref{fig2b}) also reveal transitions to bound and quasi-bound levels of H$_2^+$.
The bound states of H$_2^+$ observed in these spectra are assigned, in order of increasing energy, to the $X^+$ (18,1), (16,9), (18,3) levels, the $A^+$ (0,0), (0,2) levels, and the $X^+$ (19,1) level, the position of which is located just below the DI threshold. The broader line observed above the DI threshold is attributed to a second quasi-bound rotational level of the $X^+$ state, the (17,7) level, for which we derive by deconvolution a Lorentzian line-width function with a FWHM of 0.56(8)~cm$^{-1}$.  This conclusion is confirmed by the fact that this line appears in the H$^+$ mass channel of the MATI spectrum (see Fig.~\ref{fig2b}).

The positions of the DI threshold and of the H$_2^+$ + e$^-$ ionization thresholds gradually shift towards lower wave numbers at successive  steps of the field-ionization sequence, with shifts of $-0.68(5)$, $-0.81(5)$, $-1.07(5)$, $-1.27(5)$, and \mbox{-1.59(5)~cm$^{-1}$} for the pulses A-E. These shifts are equal to those we determine in simulations of the PFI dynamics using the method described in Ref.~\cite{hollenstein01a}. When given relative to the (17,6) and the (18,3) thresholds in Figs.~\ref{fig2a} and~\ref{fig2b}, respectively, the positions of the lines in the spectra recorded with different pulse steps are identical within the experimental uncertainties because the PFI shifts are exactly compensated. 
The positions of the levels of para and ortho H$_2^+$ determined experimentally are listed relative to the positions of the (17,6) and (18,3) levels in Table~\ref{table1}, where they are compared with the theoretical values of Moss \cite{moss93a} and the results of our own calculations. Whereas the relative positions of all levels of H$_2^+$ could be determined with uncertainties of only 0.06~cm$^{-1}$, the uncertainties in the widths of the quasi-bound levels are larger because (1) the predissociation widths of these levels are of the same magnitude as the resolution of our experiment, and (2) the DI-continuum cross section is not known and is thus difficult to cleanly remove by subtraction.
This difficulty also hindered the quantitative analysis of the (19,1) level, which is therefore omitted in Table~\ref{table1}.

The observation of transitions to states of high rotational quantum number $N^+$, up to $N^+= 9$ in ortho H$_2^+$ in Fig.~\ref{fig2b}, is attributed to the fact that, at long range, the $\bar{H}$ state has H$^+$H$^-$ ion-pair character. Consequently, a single-center expansion of the orbital out of which the electron is ejected consists of several $\ell$ components ($\ell$ is the orbital angular momentum quantum number). Applying photoionization selection rules~\cite{xie90a,signorell97a} leads to the conclusion that $|\Delta N|=|N^+-N|$ must be equal to, or less than, $\ell_{\rm max}+2 $, where $\ell_{\rm max}$ is the highest component in the single-center expansion of the $\bar{H}$ orbital, and that $\Delta N$ must be even  (odd)  for the $X^+(A^+) \leftarrow \bar{H}$ photoionizing transition. 
The spectra presented in Figs.~\ref{fig2a} and~\ref{fig2b} indicate that $\ell_{\rm max}$ is at least 4.

In Fig.~\ref{fig2}, the relative intensity of the photoelectron signal in the continuum compared to that below the DI threshold increases at each successive step of the field-ionization sequence. This trend can be explained in part by the fact that the resolution of the photoelectron spectra decreases at each step of the pulse sequence, with the consequence that sharp structures are less efficiently excited than broad ones.

\begin{table}
\centering
\caption{\small Positions of the observed (o) bound and quasi-bound states of H$_2^+$ compared with the calculated (c) values. The levels of para and ortho H$_2^+$ are given with respect to the $X^+(17,6)$ and (18,3) levels, respectively. The experimental uncertainties represent one standard deviation.}\label{table1}
\begin{ruledtabular}
\begin{tabular}{cccc}
Level & $\tilde{\nu}_\mathrm{o}(\text{cm}^{-1})$\footnotemark[1] & o$-$c($\text{cm}^{-1})$\footnotemark[2] & o$-$c($\text{cm}^{-1})$\footnotemark[1]\\
\hline
(17,6)  & 0    		& 	0 		&      0 	\\
(18,2)  & 4.349(27) 	&	0.0267      &	0.0406	\\
(0,1)    & 16.08(4) 	& 	0.0161    	&	$-$	\\
(18,4)  & 20.41(4)\footnotemark[3] 	&	$-$	&	0.0191\footnotemark[4] 	\\ \hline
(18,1)  & -14.56(4) 	&      0.0193 	&     	0.0107	\\
(16,9)  & -6.39(6) 	&      -0.0035	&     	-0.0255	\\
(18,3)  & 0			&	0		&	0		\\    
(0,0)   & 2.61(3)		&	0.0168 	& 	$-$	\\      
(0,2)   & 5.265(16)	&	0.0404   	&	$-$ 	\\    
(17,7)  & 17.11(6)\footnotemark[5]	&	$-$ 	&    -0.0105	\footnotemark[6]\\
\end{tabular}
\end{ruledtabular}
\parbox[t]{\linewidth}{\raggedright\footnotemark[1]{This work.} \footnotemark[2]{Ref.~\cite{moss93a}.} \footnotemark[3]{$\varGamma_\text{o}=0.21(7)$~cm$^{-1}$.} \footnotemark[4]{$\varGamma_\text{c}=0.20$~cm$^{-1}$.} \footnotemark[5]{$\varGamma_\text{o}=0.56(8)$~cm$^{-1}$.} \footnotemark[6]{$\varGamma_\text{c}=0.16$~cm$^{-1}$.}
}
\end{table}


Rovibrational energies $E_i$ and the nuclear wave functions $\chi_i(R)$ were calculated in atomic units by solving
\begin{equation}
\label{eq:SE}
\left[-\frac{1}{2\mu_\text{vib}}\frac{\text{d}^2}{\text{d}R^2} + U^\text{ad} + \frac{N^+(N^++1)}{2\mu_\text{rot}R^2} - E_i\right]\chi_i(R)=0, 
\end{equation}
where $U^\text{ad}=U^\text{CN}+H_1+H_2$ is the adiabatic potential curve with the clamped-nuclei energy $U^\text{CN}=U^\text{el}+1/R$ and the electronic energy $U^\text{el}$ is obtained by solving the electronic Schr\"odinger equation at fixed $R$. $U^\text{CN}$ and the adiabatic corrections $H_1=-\frac{1}{2\mu}\int{\psi_i}^*\Delta_R\psi_i \text{d}r$ and $H_2=-\frac{1}{8\mu}\int{\psi_i}^*\Delta_r\psi_i \text{d}r$ were taken from \cite{bishop73a, kolos69a}. 
Because the $X^+$ state is well separated from other \emph{gerade} states, the leading term of the nonadiabatic corrections can be evaluated conveniently by introducing $R$-dependent reduced masses for vibration and rotation, which allows one to retain the idea of a single electronic potential function \cite{kutzelnigg07a, jaquet08a}. 
Vibrational and rotational masses $\mu_\text{vib}^{-1}=\mu^{-1}\left(1+A(R)/m_\text{p} \right)$ and $\mu_\text{rot}^{-1}=\mu^{-1}\left(1+B_\text{pol}(R)/m_\text{p} \right)$ were determined using $A(R)$ and $B_\text{pol}(R)$ as given in \cite{jaquet08a}. The proton-to-electron mass ratio was taken to be $m_\text{p}/m_\text{e}=1836.15267389(17)$ and $E_\text{h}/hc=2194746.313702(13)~\text{cm}^{-1}$ \cite{mohr14a}. The relativistic and radiative corrections as reported by Moss \cite{moss93a} were added to our nonadiabatic energies. 

We implemented the renormalized Numerov method as described in \cite{johnson77a} to solve Eq.~(\ref{eq:SE}) numerically on a grid $(0.2~a_0, R_\text{max}=200~a_0)$ with an integration step of $0.01~a_0$. $U^\text{CN}$ was interpolated with a fifth-degree polynomial that fits $U^\text{el}$ and $\text{d}U^\text{el}/\text{d}R$ simultaneously at three points \cite{wolniewicz66a}. $\text{d}U^\text{el}/\text{d}R$ was calculated from $U^\text{el}$ and the adiabatic correction $H_2$ using the virial theorem which holds exactly within the Born-Oppenheimer approximation \cite{teller70a}. The other functions were interpolated using a fifth-degree Lagrange polynomial and all functions were smoothly connected to the $\text{H}^++\text{H}(1\text{s})$ dissociation limit. 
The energy $E_\text{res}$ and the FWHM $\varGamma$ of the resonances were determined by calculating the energy-dependent phase shift $\delta_{N^+}(E)$ for each $N^+$ \cite{smith71a}. Because $\lim_{R\to\infty}U^\text{ad}(R) = \text{const.}$, the asymptotic solution of Eq.~(\ref{eq:SE}) is a linear combination of the regular and irregular spherical Bessel functions $j_{N^+}(kR)$ and $n_{N^+}(kR)$ with $k^2=2\mu(E-U^\text{ad})$. The phase shift for a given energy $\delta_{N^+}(E)$ was obtained by using the values of the wave function at the two outermost grid points $R_\text{a}$ and $R_\text{b}=R_\text{max}$ using
\begin{equation}
\tan{\delta_{N^+}} = \frac{ Kj_{N^+}(R_\text{a}) - j_{N^+}(R_\text{b}) }{ Kn_{N^+}(R_\text{a}) - n_{N^+}(R_\text{b}) }; ~ K=\frac{ R_\text{a}\chi_{N^+}(R_\text{b}) }{ R_\text{b}\chi_{N^+}(R_\text{a})  }.
\end{equation}
For an isolated resonance in a single channel the energy dependence of the phase shift is given by
\begin{equation}
\tan\left[\delta_{N^+}(E) - \delta_{N^+}^0 \right] = \frac{\varGamma}{2(E_\text{res}-E)},
\end{equation}
where $\delta_{N^+}^0$ is assumed to be constant near $E_\text{res}$.

The experimental positions of bound and quasi-bound levels of H$_2^+$ agree with the calculated positions within the experimental uncertainty of ~0.06~cm$^{-1}$. The positions of the bound levels we calculate with our effective potential agree with the results of Moss within 0.025~cm$^{-1}$ and we attribute the differences to the incomplete description of the nonadiabatic effects in the present work. 
The width we observe for the (18,4) quasi-bound level agrees with the calculated value, but the measured width of the (17,7) resonance is more than three times larger than the width we calculate. Given the excellent agreement of the positions, we do not have a good explanation for this discrepancy. It may simply be a consequence of the approximate nature of our calculations. Alternatively the ions generated in the DI continuum of ortho H$_2$, which is more than three times stronger than in para H$_2$, may broaden the PFI-ZEKE signal. The discrepancy may further indicate nonadiabatic interactions in the three-body system H(1s)\--e$^-$\--H$^+$ which, in this energy region, may decay either by ionization or dissociation with chaotic branching ratios. The fact that the field-ionization shifts behave normally speaks against the latter two explanations. Further theoretical work is needed to clarify this discrepancy.

We thank Dr. Ch. Jungen (Orsay) for useful discussions.
The content of this letter is related to material presented in May 2015 during the Kolos Lecture at the Department of Chemistry, University of Warsaw.
This work is supported financially by the Swiss National Science Foundation under project SNF 200020-159848.


\begin{thebibliography}{65}
\makeatletter
\providecommand \@ifxundefined [1]{%
 \@ifx{#1\undefined}
}%
\providecommand \@ifnum [1]{%
 \ifnum #1\expandafter \@firstoftwo
 \else \expandafter \@secondoftwo
 \fi
}%
\providecommand \@ifx [1]{%
 \ifx #1\expandafter \@firstoftwo
 \else \expandafter \@secondoftwo
 \fi
}%
\providecommand \natexlab [1]{#1}%
\providecommand \enquote  [1]{``#1''}%
\providecommand \bibnamefont  [1]{#1}%
\providecommand \bibfnamefont [1]{#1}%
\providecommand \citenamefont [1]{#1}%
\providecommand \href@noop [0]{\@secondoftwo}%
\providecommand \href [0]{\begingroup \@sanitize@url \@href}%
\providecommand \@href[1]{\@@startlink{#1}\@@href}%
\providecommand \@@href[1]{\endgroup#1\@@endlink}%
\providecommand \@sanitize@url [0]{\catcode `\\12\catcode `\$12\catcode
  `\&12\catcode `\#12\catcode `\^12\catcode `\_12\catcode `\%12\relax}%
\providecommand \@@startlink[1]{}%
\providecommand \@@endlink[0]{}%
\providecommand \url  [0]{\begingroup\@sanitize@url \@url }%
\providecommand \@url [1]{\endgroup\@href {#1}{\urlprefix }}%
\providecommand \urlprefix  [0]{URL }%
\providecommand \Eprint [0]{\href }%
\providecommand \doibase [0]{http://dx.doi.org/}%
\providecommand \selectlanguage [0]{\@gobble}%
\providecommand \bibinfo  [0]{\@secondoftwo}%
\providecommand \bibfield  [0]{\@secondoftwo}%
\providecommand \translation [1]{[#1]}%
\providecommand \BibitemOpen [0]{}%
\providecommand \bibitemStop [0]{}%
\providecommand \bibitemNoStop [0]{.\EOS\space}%
\providecommand \EOS [0]{\spacefactor3000\relax}%
\providecommand \BibitemShut  [1]{\csname bibitem#1\endcsname}%
\let\auto@bib@innerbib\@empty
\bibitem [{\citenamefont {Kolos}\ and\ \citenamefont
  {Wolniewicz}(1963)}]{kolos63a}%
  \BibitemOpen
  \bibfield  {author} {\bibinfo {author} {\bibfnamefont {W.}~\bibnamefont
  {Kolos}}\ and\ \bibinfo {author} {\bibfnamefont {L.}~\bibnamefont
  {Wolniewicz}},\ }\href {\doibase 10.1103/RevModPhys.35.473} {\bibfield
  {journal} {\bibinfo  {journal} {Rev. Mod. Phys.}\ }\textbf {\bibinfo {volume}
  {35}},\ \bibinfo {pages} {473} (\bibinfo {year} {1963})}\BibitemShut
  {NoStop}%
\bibitem [{\citenamefont {Kolos}\ and\ \citenamefont
  {Rychlewski}(1993)}]{kolos93a}%
  \BibitemOpen
  \bibfield  {author} {\bibinfo {author} {\bibfnamefont {W.}~\bibnamefont
  {Kolos}}\ and\ \bibinfo {author} {\bibfnamefont {J.}~\bibnamefont
  {Rychlewski}},\ }\href {\doibase 10.1063/1.464023} {\bibfield  {journal}
  {\bibinfo  {journal} {J. Chem. Phys.}\ }\textbf {\bibinfo {volume} {98}},\
  \bibinfo {pages} {3960} (\bibinfo {year} {1993})}\BibitemShut {NoStop}%
\bibitem [{\citenamefont {Wolniewicz}(1993)}]{wolniewicz93a}%
  \BibitemOpen
  \bibfield  {author} {\bibinfo {author} {\bibfnamefont {L.}~\bibnamefont
  {Wolniewicz}},\ }\href {\doibase 10.1063/1.465303} {\bibfield  {journal}
  {\bibinfo  {journal} {J. Chem. Phys.}\ }\textbf {\bibinfo {volume} {99}},\
  \bibinfo {pages} {1851} (\bibinfo {year} {1993})}\BibitemShut {NoStop}%
\bibitem [{\citenamefont {Kolos}(1994)}]{kolos94a}%
  \BibitemOpen
  \bibfield  {author} {\bibinfo {author} {\bibfnamefont {W.}~\bibnamefont
  {Kolos}},\ }\href {\doibase 10.1063/1.467825} {\bibfield  {journal} {\bibinfo
   {journal} {J. Chem. Phys.}\ }\textbf {\bibinfo {volume} {101}},\ \bibinfo
  {pages} {1330} (\bibinfo {year} {1994})}\BibitemShut {NoStop}%
\bibitem [{\citenamefont {Wolniewicz}(1995)}]{wolniewicz95a}%
  \BibitemOpen
  \bibfield  {author} {\bibinfo {author} {\bibfnamefont {L.}~\bibnamefont
  {Wolniewicz}},\ }\href {\doibase 10.1006/jmsp.1995.1027} {\bibfield
  {journal} {\bibinfo  {journal} {J. Mol. Spectrosc.}\ }\textbf {\bibinfo
  {volume} {169}},\ \bibinfo {pages} {329} (\bibinfo {year}
  {1995})}\BibitemShut {NoStop}%
\bibitem [{\citenamefont {Piszczatowski}\ \emph {et~al.}(2009)\citenamefont
  {Piszczatowski}, \citenamefont {{\L}ach}, \citenamefont {Przybytek},
  \citenamefont {Komasa}, \citenamefont {Pachucki},\ and\ \citenamefont
  {Jeziorski}}]{piszczatowski09a}%
  \BibitemOpen
  \bibfield  {author} {\bibinfo {author} {\bibfnamefont {K.}~\bibnamefont
  {Piszczatowski}}, \bibinfo {author} {\bibfnamefont {G.}~\bibnamefont
  {{\L}ach}}, \bibinfo {author} {\bibfnamefont {M.}~\bibnamefont {Przybytek}},
  \bibinfo {author} {\bibfnamefont {J.}~\bibnamefont {Komasa}}, \bibinfo
  {author} {\bibfnamefont {K.}~\bibnamefont {Pachucki}}, \ and\ \bibinfo
  {author} {\bibfnamefont {B.}~\bibnamefont {Jeziorski}},\ }\href {\doibase
  10.1021/ct900391p} {\bibfield  {journal} {\bibinfo  {journal} {J. Chem.
  Theory Comput.}\ }\textbf {\bibinfo {volume} {5}},\ \bibinfo {pages} {3039}
  (\bibinfo {year} {2009})}\BibitemShut {NoStop}%
\bibitem [{\citenamefont {Bishop}\ and\ \citenamefont
  {Wetmore}(1973)}]{bishop73a}%
  \BibitemOpen
  \bibfield  {author} {\bibinfo {author} {\bibfnamefont {D.~M.}\ \bibnamefont
  {Bishop}}\ and\ \bibinfo {author} {\bibfnamefont {R.~W.}\ \bibnamefont
  {Wetmore}},\ }\href {\doibase 10.1080/00268977300101461} {\bibfield
  {journal} {\bibinfo  {journal} {Mol. Phys.}\ }\textbf {\bibinfo {volume}
  {26}},\ \bibinfo {pages} {145} (\bibinfo {year} {1973})}\BibitemShut
  {NoStop}%
\bibitem [{\citenamefont {Bishop}(1977)}]{bishop77a}%
  \BibitemOpen
  \bibfield  {author} {\bibinfo {author} {\bibfnamefont {D.~M.}\ \bibnamefont
  {Bishop}},\ }\href {\doibase 10.1063/1.434382} {\bibfield  {journal}
  {\bibinfo  {journal} {J. Chem. Phys.}\ }\textbf {\bibinfo {volume} {66}},\
  \bibinfo {pages} {3842} (\bibinfo {year} {1977})}\BibitemShut {NoStop}%
\bibitem [{\citenamefont {Bishop}\ and\ \citenamefont
  {Cheung}(1978)}]{bishop78a}%
  \BibitemOpen
  \bibfield  {author} {\bibinfo {author} {\bibfnamefont {D.~M.}\ \bibnamefont
  {Bishop}}\ and\ \bibinfo {author} {\bibfnamefont {L.~M.}\ \bibnamefont
  {Cheung}},\ }\href {\doibase 10.1088/0022-3700/11/18/010} {\bibfield
  {journal} {\bibinfo  {journal} {J. Phys. B}\ }\textbf {\bibinfo {volume}
  {11}},\ \bibinfo {pages} {3133} (\bibinfo {year} {1978})}\BibitemShut
  {NoStop}%
\bibitem [{\citenamefont {Wolniewicz}\ and\ \citenamefont
  {Poll}(1986)}]{wolniewicz86a}%
  \BibitemOpen
  \bibfield  {author} {\bibinfo {author} {\bibfnamefont {L.}~\bibnamefont
  {Wolniewicz}}\ and\ \bibinfo {author} {\bibfnamefont {J.~D.}\ \bibnamefont
  {Poll}},\ }\href {\doibase 10.1080/00268978600102501} {\bibfield  {journal}
  {\bibinfo  {journal} {Mol. Phys.}\ }\textbf {\bibinfo {volume} {59}},\
  \bibinfo {pages} {953} (\bibinfo {year} {1986})}\BibitemShut {NoStop}%
\bibitem [{\citenamefont {Bunker}\ and\ \citenamefont
  {Moss}(1977)}]{bunker77a}%
  \BibitemOpen
  \bibfield  {author} {\bibinfo {author} {\bibfnamefont {P.~R.}\ \bibnamefont
  {Bunker}}\ and\ \bibinfo {author} {\bibfnamefont {R.~E.}\ \bibnamefont
  {Moss}},\ }\href {\doibase 10.1080/00268977700100351} {\bibfield  {journal}
  {\bibinfo  {journal} {Mol. Phys.}\ }\textbf {\bibinfo {volume} {33}},\
  \bibinfo {pages} {417} (\bibinfo {year} {1977})}\BibitemShut {NoStop}%
\bibitem [{\citenamefont {Moss}(1993)}]{moss93a}%
  \BibitemOpen
  \bibfield  {author} {\bibinfo {author} {\bibfnamefont {R.~E.}\ \bibnamefont
  {Moss}},\ }\href {\doibase 10.1080/00268979300103211} {\bibfield  {journal}
  {\bibinfo  {journal} {Mol. Phys.}\ }\textbf {\bibinfo {volume} {80}},\
  \bibinfo {pages} {1541} (\bibinfo {year} {1993})}\BibitemShut {NoStop}%
\bibitem [{\citenamefont {Herman}\ and\ \citenamefont
  {Asgharian}(1966)}]{herman66a}%
  \BibitemOpen
  \bibfield  {author} {\bibinfo {author} {\bibfnamefont {R.~M.}\ \bibnamefont
  {Herman}}\ and\ \bibinfo {author} {\bibfnamefont {A.}~\bibnamefont
  {Asgharian}},\ }\href {\doibase 10.1016/0022-2852(66)90254-2} {\bibfield
  {journal} {\bibinfo  {journal} {J. Mol. Spectrosc.}\ }\textbf {\bibinfo
  {volume} {19}},\ \bibinfo {pages} {305} (\bibinfo {year} {1966})}\BibitemShut
  {NoStop}%
\bibitem [{\citenamefont {Schwenke}(2001)}]{schwenke01a}%
  \BibitemOpen
  \bibfield  {author} {\bibinfo {author} {\bibfnamefont {D.~W.}\ \bibnamefont
  {Schwenke}},\ }\href {\doibase 10.1063/1.1334897} {\bibfield  {journal}
  {\bibinfo  {journal} {J. Chem. Phys.}\ }\textbf {\bibinfo {volume} {114}},\
  \bibinfo {pages} {1693} (\bibinfo {year} {2001})}\BibitemShut {NoStop}%
\bibitem [{\citenamefont {Kutzelnigg}(2007)}]{kutzelnigg07a}%
  \BibitemOpen
  \bibfield  {author} {\bibinfo {author} {\bibfnamefont {W.}~\bibnamefont
  {Kutzelnigg}},\ }\href {\doibase 10.1080/00268970701604671} {\bibfield
  {journal} {\bibinfo  {journal} {Mol. Phys.}\ }\textbf {\bibinfo {volume}
  {105}},\ \bibinfo {pages} {2627} (\bibinfo {year} {2007})}\BibitemShut
  {NoStop}%
\bibitem [{\citenamefont {Jaquet}\ and\ \citenamefont
  {Kutzelnigg}(2008)}]{jaquet08a}%
  \BibitemOpen
  \bibfield  {author} {\bibinfo {author} {\bibfnamefont {R.}~\bibnamefont
  {Jaquet}}\ and\ \bibinfo {author} {\bibfnamefont {W.}~\bibnamefont
  {Kutzelnigg}},\ }\href {\doibase 10.1016/j.chemphys.2008.02.068} {\bibfield
  {journal} {\bibinfo  {journal} {Chem. Phys.}\ }\textbf {\bibinfo {volume}
  {346}},\ \bibinfo {pages} {69} (\bibinfo {year} {2008})}\BibitemShut
  {NoStop}%
\bibitem [{\citenamefont {Diniz}\ \emph {et~al.}(2012)\citenamefont {Diniz},
  \citenamefont {Alijah},\ and\ \citenamefont {Mohallem}}]{diniz12a}%
  \BibitemOpen
  \bibfield  {author} {\bibinfo {author} {\bibfnamefont {L.~G.}\ \bibnamefont
  {Diniz}}, \bibinfo {author} {\bibfnamefont {A.}~\bibnamefont {Alijah}}, \
  and\ \bibinfo {author} {\bibfnamefont {J.~R.}\ \bibnamefont {Mohallem}},\
  }\href {\doibase 10.1063/1.4762442} {\bibfield  {journal} {\bibinfo
  {journal} {J. Chem. Phys.}\ }\textbf {\bibinfo {volume} {137}},\ \bibinfo
  {pages} {164316} (\bibinfo {year} {2012})}\BibitemShut {NoStop}%
\bibitem [{\citenamefont {Taylor}\ \emph {et~al.}(1999)\citenamefont {Taylor},
  \citenamefont {Yan}, \citenamefont {Dalgarno},\ and\ \citenamefont
  {Babb}}]{taylor99a}%
  \BibitemOpen
  \bibfield  {author} {\bibinfo {author} {\bibfnamefont {J.~M.}\ \bibnamefont
  {Taylor}}, \bibinfo {author} {\bibfnamefont {Z.-C.}\ \bibnamefont {Yan}},
  \bibinfo {author} {\bibfnamefont {A.}~\bibnamefont {Dalgarno}}, \ and\
  \bibinfo {author} {\bibfnamefont {J.~F.}\ \bibnamefont {Babb}},\ }\href
  {\doibase 10.1080/00268979909482806} {\bibfield  {journal} {\bibinfo
  {journal} {Mol. Phys.}\ }\textbf {\bibinfo {volume} {97}},\ \bibinfo {pages}
  {25} (\bibinfo {year} {1999})}\BibitemShut {NoStop}%
\bibitem [{\citenamefont {Korobov}(2000)}]{korobov00a}%
  \BibitemOpen
  \bibfield  {author} {\bibinfo {author} {\bibfnamefont {V.~I.}\ \bibnamefont
  {Korobov}},\ }\href {\doibase 10.1103/PhysRevA.61.064503} {\bibfield
  {journal} {\bibinfo  {journal} {Phys. Rev. A}\ }\textbf {\bibinfo {volume}
  {61}},\ \bibinfo {pages} {064503} (\bibinfo {year} {2000})}\BibitemShut
  {NoStop}%
\bibitem [{\citenamefont {Hilico}\ \emph {et~al.}(2000)\citenamefont {Hilico},
  \citenamefont {Billy}, \citenamefont {Gr{\'e}maud},\ and\ \citenamefont
  {Delande}}]{hilico00a}%
  \BibitemOpen
  \bibfield  {author} {\bibinfo {author} {\bibfnamefont {L.}~\bibnamefont
  {Hilico}}, \bibinfo {author} {\bibfnamefont {N.}~\bibnamefont {Billy}},
  \bibinfo {author} {\bibfnamefont {B.}~\bibnamefont {Gr{\'e}maud}}, \ and\
  \bibinfo {author} {\bibfnamefont {D.}~\bibnamefont {Delande}},\ }\href
  {\doibase 10.1007/s100530070006} {\bibfield  {journal} {\bibinfo  {journal}
  {Eur. Phys. J. D}\ }\textbf {\bibinfo {volume} {12}},\ \bibinfo {pages} {449}
  (\bibinfo {year} {2000})}\BibitemShut {NoStop}%
\bibitem [{\citenamefont {M{\'a}tyus}\ and\ \citenamefont
  {Reiher}(2012)}]{matyus12a}%
  \BibitemOpen
  \bibfield  {author} {\bibinfo {author} {\bibfnamefont {E.}~\bibnamefont
  {M{\'a}tyus}}\ and\ \bibinfo {author} {\bibfnamefont {M.}~\bibnamefont
  {Reiher}},\ }\href {\doibase 10.1063/1.4731696} {\bibfield  {journal}
  {\bibinfo  {journal} {J. Chem. Phys.}\ }\textbf {\bibinfo {volume} {137}},\
  \bibinfo {pages} {024104} (\bibinfo {year} {2012})}\BibitemShut {NoStop}%
\bibitem [{\citenamefont {Stanke}\ and\ \citenamefont
  {Adamowicz}(2013)}]{stanke13a}%
  \BibitemOpen
  \bibfield  {author} {\bibinfo {author} {\bibfnamefont {M.}~\bibnamefont
  {Stanke}}\ and\ \bibinfo {author} {\bibfnamefont {L.}~\bibnamefont
  {Adamowicz}},\ }\href {\doibase 10.1021/jp4020492} {\bibfield  {journal}
  {\bibinfo  {journal} {J. Phys. Chem. A}\ }\textbf {\bibinfo {volume} {117}},\
  \bibinfo {pages} {10129} (\bibinfo {year} {2013})}\BibitemShut {NoStop}%
\bibitem [{\citenamefont {Bukowski}\ \emph {et~al.}(1992)\citenamefont
  {Bukowski}, \citenamefont {Jeziorski}, \citenamefont {Moszy{\'n}ski},\ and\
  \citenamefont {Ko{\l}os}}]{bukowski92a}%
  \BibitemOpen
  \bibfield  {author} {\bibinfo {author} {\bibfnamefont {R.}~\bibnamefont
  {Bukowski}}, \bibinfo {author} {\bibfnamefont {B.}~\bibnamefont {Jeziorski}},
  \bibinfo {author} {\bibfnamefont {R.}~\bibnamefont {Moszy{\'n}ski}}, \ and\
  \bibinfo {author} {\bibfnamefont {W.}~\bibnamefont {Ko{\l}os}},\ }\href
  {\doibase 10.1002/qua.560420205} {\bibfield  {journal} {\bibinfo  {journal}
  {Int. J. quant. Chem.}\ }\textbf {\bibinfo {volume} {42}},\ \bibinfo {pages}
  {287} (\bibinfo {year} {1992})}\BibitemShut {NoStop}%
\bibitem [{\citenamefont {Korobov}(2006)}]{korobov06a}%
  \BibitemOpen
  \bibfield  {author} {\bibinfo {author} {\bibfnamefont {V.~I.}\ \bibnamefont
  {Korobov}},\ }\href {\doibase 10.1103/PhysRevA.74.052506} {\bibfield
  {journal} {\bibinfo  {journal} {Phys. Rev. A}\ }\textbf {\bibinfo {volume}
  {74}},\ \bibinfo {pages} {052506} (\bibinfo {year} {2006})}\BibitemShut
  {NoStop}%
\bibitem [{\citenamefont {Korobov}(2008)}]{korobov08a}%
  \BibitemOpen
  \bibfield  {author} {\bibinfo {author} {\bibfnamefont {V.~I.}\ \bibnamefont
  {Korobov}},\ }\href {\doibase 10.1103/PhysRevA.77.022509} {\bibfield
  {journal} {\bibinfo  {journal} {Phys. Rev. A}\ }\textbf {\bibinfo {volume}
  {77}},\ \bibinfo {pages} {022509} (\bibinfo {year} {2008})}\BibitemShut
  {NoStop}%
\bibitem [{\citenamefont {Korobov}\ \emph {et~al.}(2014)\citenamefont
  {Korobov}, \citenamefont {Hilico},\ and\ \citenamefont {Karr}}]{korobov14a}%
  \BibitemOpen
  \bibfield  {author} {\bibinfo {author} {\bibfnamefont {V.~I.}\ \bibnamefont
  {Korobov}}, \bibinfo {author} {\bibfnamefont {L.}~\bibnamefont {Hilico}}, \
  and\ \bibinfo {author} {\bibfnamefont {J.-P.}\ \bibnamefont {Karr}},\ }\href
  {\doibase 10.1103/PhysRevA.89.032511} {\bibfield  {journal} {\bibinfo
  {journal} {Phys. Rev. A}\ }\textbf {\bibinfo {volume} {89}},\ \bibinfo
  {pages} {032511} (\bibinfo {year} {2014})}\BibitemShut {NoStop}%
\bibitem [{\citenamefont {Komasa}\ \emph {et~al.}(2011)\citenamefont {Komasa},
  \citenamefont {Piszczatowski}, \citenamefont {{\L}ach}, \citenamefont
  {Przybytek}, \citenamefont {Jeziorski},\ and\ \citenamefont
  {Pachucki}}]{komasa11a}%
  \BibitemOpen
  \bibfield  {author} {\bibinfo {author} {\bibfnamefont {J.}~\bibnamefont
  {Komasa}}, \bibinfo {author} {\bibfnamefont {K.}~\bibnamefont
  {Piszczatowski}}, \bibinfo {author} {\bibfnamefont {G.}~\bibnamefont
  {{\L}ach}}, \bibinfo {author} {\bibfnamefont {M.}~\bibnamefont {Przybytek}},
  \bibinfo {author} {\bibfnamefont {B.}~\bibnamefont {Jeziorski}}, \ and\
  \bibinfo {author} {\bibfnamefont {K.}~\bibnamefont {Pachucki}},\ }\href
  {\doibase 10.1021/ct200438t} {\bibfield  {journal} {\bibinfo  {journal} {J.
  Chem. Theory Comput.}\ }\textbf {\bibinfo {volume} {7}},\ \bibinfo {pages}
  {3105} (\bibinfo {year} {2011})}\BibitemShut {NoStop}%
\bibitem [{\citenamefont {Ehrenfreund}\ and\ \citenamefont
  {Charnley}(2000)}]{ehrenfreund00a}%
  \BibitemOpen
  \bibfield  {author} {\bibinfo {author} {\bibfnamefont {P.}~\bibnamefont
  {Ehrenfreund}}\ and\ \bibinfo {author} {\bibfnamefont {S.~B.}\ \bibnamefont
  {Charnley}},\ }\href {\doibase 10.1146/annurev.astro.38.1.427} {\bibfield
  {journal} {\bibinfo  {journal} {Annu. Rev. Astron. Astrophys.}\ }\textbf
  {\bibinfo {volume} {38}},\ \bibinfo {pages} {427} (\bibinfo {year}
  {2000})}\BibitemShut {NoStop}%
\bibitem [{\citenamefont {Hirata}\ and\ \citenamefont
  {Padmanabhan}(2006)}]{hirata06a}%
  \BibitemOpen
  \bibfield  {author} {\bibinfo {author} {\bibfnamefont {C.~M.}\ \bibnamefont
  {Hirata}}\ and\ \bibinfo {author} {\bibfnamefont {N.}~\bibnamefont
  {Padmanabhan}},\ }\href {\doibase 10.1111/j.1365-2966.2006.10924.x}
  {\bibfield  {journal} {\bibinfo  {journal} {Mon. Not. R. Astron. Soc.}\
  }\textbf {\bibinfo {volume} {372}},\ \bibinfo {pages} {1175} (\bibinfo {year}
  {2006})}\BibitemShut {NoStop}%
\bibitem [{\citenamefont {Peek}(1969)}]{peek69a}%
  \BibitemOpen
  \bibfield  {author} {\bibinfo {author} {\bibfnamefont {J.~M.}\ \bibnamefont
  {Peek}},\ }\href {\doibase 10.1063/1.1670939} {\bibfield  {journal} {\bibinfo
   {journal} {J. Chem. Phys.}\ }\textbf {\bibinfo {volume} {50}},\ \bibinfo
  {pages} {4595} (\bibinfo {year} {1969})}\BibitemShut {NoStop}%
\bibitem [{\citenamefont {Carbonell}\ \emph {et~al.}(2003)\citenamefont
  {Carbonell}, \citenamefont {Lazauskas}, \citenamefont {Delande},
  \citenamefont {Hilico},\ and\ \citenamefont {Kili{\c c}}}]{carbonell03a}%
  \BibitemOpen
  \bibfield  {author} {\bibinfo {author} {\bibfnamefont {J.}~\bibnamefont
  {Carbonell}}, \bibinfo {author} {\bibfnamefont {R.}~\bibnamefont
  {Lazauskas}}, \bibinfo {author} {\bibfnamefont {D.}~\bibnamefont {Delande}},
  \bibinfo {author} {\bibfnamefont {L.}~\bibnamefont {Hilico}}, \ and\ \bibinfo
  {author} {\bibfnamefont {S.}~\bibnamefont {Kili{\c c}}},\ }\href {\doibase
  10.1209/epl/i2003-00176-1} {\bibfield  {journal} {\bibinfo  {journal}
  {Europhys. Lett.}\ }\textbf {\bibinfo {volume} {64}},\ \bibinfo {pages} {316}
  (\bibinfo {year} {2003})}\BibitemShut {NoStop}%
\bibitem [{\citenamefont {Carbonell}\ \emph {et~al.}(2004)\citenamefont
  {Carbonell}, \citenamefont {Lazauskas},\ and\ \citenamefont
  {Korobov}}]{carbonell04a}%
  \BibitemOpen
  \bibfield  {author} {\bibinfo {author} {\bibfnamefont {J.}~\bibnamefont
  {Carbonell}}, \bibinfo {author} {\bibfnamefont {R.}~\bibnamefont
  {Lazauskas}}, \ and\ \bibinfo {author} {\bibfnamefont {V.~I.}\ \bibnamefont
  {Korobov}},\ }\href {\doibase 10.1088/0953-4075/37/14/012} {\bibfield
  {journal} {\bibinfo  {journal} {J. Phys. B}\ }\textbf {\bibinfo {volume}
  {37}},\ \bibinfo {pages} {2997} (\bibinfo {year} {2004})}\BibitemShut
  {NoStop}%
\bibitem [{\citenamefont {Carrington}\ \emph
  {et~al.}(1989{\natexlab{a}})\citenamefont {Carrington}, \citenamefont
  {McNab},\ and\ \citenamefont {Montgomerie}}]{carrington89a}%
  \BibitemOpen
  \bibfield  {author} {\bibinfo {author} {\bibfnamefont {A.}~\bibnamefont
  {Carrington}}, \bibinfo {author} {\bibfnamefont {I.~R.}\ \bibnamefont
  {McNab}}, \ and\ \bibinfo {author} {\bibfnamefont {C.~A.}\ \bibnamefont
  {Montgomerie}},\ }\href {\doibase 10.1088/0953-4075/22/22/006} {\bibfield
  {journal} {\bibinfo  {journal} {J. Phys. B}\ }\textbf {\bibinfo {volume}
  {22}},\ \bibinfo {pages} {3551} (\bibinfo {year}
  {1989}{\natexlab{a}})}\BibitemShut {NoStop}%
\bibitem [{\citenamefont {Carrington}\ \emph
  {et~al.}(1989{\natexlab{b}})\citenamefont {Carrington}, \citenamefont
  {McNab},\ and\ \citenamefont {Montgomerie}}]{carrington89b}%
  \BibitemOpen
  \bibfield  {author} {\bibinfo {author} {\bibfnamefont {A.}~\bibnamefont
  {Carrington}}, \bibinfo {author} {\bibfnamefont {I.~R.}\ \bibnamefont
  {McNab}}, \ and\ \bibinfo {author} {\bibfnamefont {C.~A.}\ \bibnamefont
  {Montgomerie}},\ }\href {\doibase 10.1016/0009-2614(89)87589-X} {\bibfield
  {journal} {\bibinfo  {journal} {Chem. Phys. Lett.}\ }\textbf {\bibinfo
  {volume} {160}},\ \bibinfo {pages} {237} (\bibinfo {year}
  {1989}{\natexlab{b}})}\BibitemShut {NoStop}%
\bibitem [{\citenamefont {Carrington}\ \emph {et~al.}(1993)\citenamefont
  {Carrington}, \citenamefont {Leach}, \citenamefont {Moss}, \citenamefont
  {Steimle}, \citenamefont {Viant},\ and\ \citenamefont
  {West}}]{carrington93a}%
  \BibitemOpen
  \bibfield  {author} {\bibinfo {author} {\bibfnamefont {A.}~\bibnamefont
  {Carrington}}, \bibinfo {author} {\bibfnamefont {C.~A.}\ \bibnamefont
  {Leach}}, \bibinfo {author} {\bibfnamefont {R.~E.}\ \bibnamefont {Moss}},
  \bibinfo {author} {\bibfnamefont {T.~C.}\ \bibnamefont {Steimle}}, \bibinfo
  {author} {\bibfnamefont {M.~R.}\ \bibnamefont {Viant}}, \ and\ \bibinfo
  {author} {\bibfnamefont {Y.~D.}\ \bibnamefont {West}},\ }\href {\doibase
  10.1039/FT9938900603} {\bibfield  {journal} {\bibinfo  {journal} {J. Chem.
  Soc., Faraday Trans.}\ }\textbf {\bibinfo {volume} {89}},\ \bibinfo {pages}
  {603} (\bibinfo {year} {1993})}\BibitemShut {NoStop}%
\bibitem [{\citenamefont {Critchley}\ \emph {et~al.}(2001)\citenamefont
  {Critchley}, \citenamefont {Hughes},\ and\ \citenamefont
  {McNab}}]{critchley01a}%
  \BibitemOpen
  \bibfield  {author} {\bibinfo {author} {\bibfnamefont {A.~D.~J.}\
  \bibnamefont {Critchley}}, \bibinfo {author} {\bibfnamefont {A.~N.}\
  \bibnamefont {Hughes}}, \ and\ \bibinfo {author} {\bibfnamefont {I.~R.}\
  \bibnamefont {McNab}},\ }\href {\doibase 10.1103/PhysRevLett.86.1725}
  {\bibfield  {journal} {\bibinfo  {journal} {Phys. Rev. Lett.}\ }\textbf
  {\bibinfo {volume} {86}},\ \bibinfo {pages} {1725} (\bibinfo {year}
  {2001})}\BibitemShut {NoStop}%
\bibitem [{\citenamefont {Jefferts}(1968)}]{jefferts68a}%
  \BibitemOpen
  \bibfield  {author} {\bibinfo {author} {\bibfnamefont {K.~B.}\ \bibnamefont
  {Jefferts}},\ }\href {\doibase 10.1103/PhysRevLett.20.39} {\bibfield
  {journal} {\bibinfo  {journal} {Phys. Rev. Lett.}\ }\textbf {\bibinfo
  {volume} {20}},\ \bibinfo {pages} {39} (\bibinfo {year} {1968})}\BibitemShut
  {NoStop}%
\bibitem [{\citenamefont {Jefferts}(1969)}]{jefferts69a}%
  \BibitemOpen
  \bibfield  {author} {\bibinfo {author} {\bibfnamefont {K.~B.}\ \bibnamefont
  {Jefferts}},\ }\href {\doibase 10.1103/PhysRevLett.23.1476} {\bibfield
  {journal} {\bibinfo  {journal} {Phys. Rev. Lett.}\ }\textbf {\bibinfo
  {volume} {23}},\ \bibinfo {pages} {1476} (\bibinfo {year}
  {1969})}\BibitemShut {NoStop}%
\bibitem [{\citenamefont {{\AA}sbrink}(1970)}]{asbrink70a}%
  \BibitemOpen
  \bibfield  {author} {\bibinfo {author} {\bibfnamefont {L.}~\bibnamefont
  {{\AA}sbrink}},\ }\href {\doibase 10.1016/0009-2614(70)80169-5} {\bibfield
  {journal} {\bibinfo  {journal} {Chem. Phys. Lett.}\ }\textbf {\bibinfo
  {volume} {7}},\ \bibinfo {pages} {549} (\bibinfo {year} {1970})}\BibitemShut
  {NoStop}%
\bibitem [{\citenamefont {Merkt}\ and\ \citenamefont
  {Softley}(1992)}]{merkt92a}%
  \BibitemOpen
  \bibfield  {author} {\bibinfo {author} {\bibfnamefont {F.}~\bibnamefont
  {Merkt}}\ and\ \bibinfo {author} {\bibfnamefont {T.~P.}\ \bibnamefont
  {Softley}},\ }\href {\doibase 10.1063/1.461870} {\bibfield  {journal}
  {\bibinfo  {journal} {J. Chem. Phys.}\ }\textbf {\bibinfo {volume} {96}},\
  \bibinfo {pages} {4149} (\bibinfo {year} {1992})}\BibitemShut {NoStop}%
\bibitem [{\citenamefont {Chang}\ \emph {et~al.}(2007)\citenamefont {Chang},
  \citenamefont {Ng}, \citenamefont {Stimson}, \citenamefont {Evans},\ and\
  \citenamefont {Hsu}}]{chang07a}%
  \BibitemOpen
  \bibfield  {author} {\bibinfo {author} {\bibfnamefont {C.}~\bibnamefont
  {Chang}}, \bibinfo {author} {\bibfnamefont {C.-Y.}\ \bibnamefont {Ng}},
  \bibinfo {author} {\bibfnamefont {S.}~\bibnamefont {Stimson}}, \bibinfo
  {author} {\bibfnamefont {M.}~\bibnamefont {Evans}}, \ and\ \bibinfo {author}
  {\bibfnamefont {C.~W.}\ \bibnamefont {Hsu}},\ }\href {\doibase
  10.1088/1674-0068/20/04/352-364} {\bibfield  {journal} {\bibinfo  {journal}
  {Chin. J. Chem. Phys.}\ }\textbf {\bibinfo {volume} {20}},\ \bibinfo {pages}
  {352} (\bibinfo {year} {2007})}\BibitemShut {NoStop}%
\bibitem [{\citenamefont {Herzberg}\ and\ \citenamefont
  {Jungen}(1972)}]{herzberg72a}%
  \BibitemOpen
  \bibfield  {author} {\bibinfo {author} {\bibfnamefont {G.}~\bibnamefont
  {Herzberg}}\ and\ \bibinfo {author} {\bibfnamefont {C.}~\bibnamefont
  {Jungen}},\ }\href {\doibase 10.1016/0022-2852(72)90064-1} {\bibfield
  {journal} {\bibinfo  {journal} {J. Mol. Spectrosc.}\ }\textbf {\bibinfo
  {volume} {41}},\ \bibinfo {pages} {425} (\bibinfo {year} {1972})}\BibitemShut
  {NoStop}%
\bibitem [{\citenamefont {Arcuni}\ \emph {et~al.}(1990)\citenamefont {Arcuni},
  \citenamefont {Hessels},\ and\ \citenamefont {Lundeen}}]{arcuni90a}%
  \BibitemOpen
  \bibfield  {author} {\bibinfo {author} {\bibfnamefont {P.~W.}\ \bibnamefont
  {Arcuni}}, \bibinfo {author} {\bibfnamefont {E.~A.}\ \bibnamefont {Hessels}},
  \ and\ \bibinfo {author} {\bibfnamefont {S.~R.}\ \bibnamefont {Lundeen}},\
  }\href {\doibase 10.1103/PhysRevA.41.3648} {\bibfield  {journal} {\bibinfo
  {journal} {Phys. Rev. A}\ }\textbf {\bibinfo {volume} {41}},\ \bibinfo
  {pages} {3648} (\bibinfo {year} {1990})}\BibitemShut {NoStop}%
\bibitem [{\citenamefont {Osterwalder}\ \emph {et~al.}(2004)\citenamefont
  {Osterwalder}, \citenamefont {W{\"u}est}, \citenamefont {Merkt},\ and\
  \citenamefont {Jungen}}]{osterwalder04a}%
  \BibitemOpen
  \bibfield  {author} {\bibinfo {author} {\bibfnamefont {A.}~\bibnamefont
  {Osterwalder}}, \bibinfo {author} {\bibfnamefont {A.}~\bibnamefont
  {W{\"u}est}}, \bibinfo {author} {\bibfnamefont {F.}~\bibnamefont {Merkt}}, \
  and\ \bibinfo {author} {\bibfnamefont {C.}~\bibnamefont {Jungen}},\ }\href
  {\doibase 10.1063/1.1792596} {\bibfield  {journal} {\bibinfo  {journal} {J.
  Chem. Phys.}\ }\textbf {\bibinfo {volume} {121}},\ \bibinfo {pages} {11810}
  (\bibinfo {year} {2004})}\BibitemShut {NoStop}%
\bibitem [{\citenamefont {Liu}\ \emph {et~al.}(2009)\citenamefont {Liu},
  \citenamefont {Salumbides}, \citenamefont {Hollenstein}, \citenamefont
  {Koelemeij}, \citenamefont {Eikema}, \citenamefont {Ubachs},\ and\
  \citenamefont {Merkt}}]{liu09a}%
  \BibitemOpen
  \bibfield  {author} {\bibinfo {author} {\bibfnamefont {J.}~\bibnamefont
  {Liu}}, \bibinfo {author} {\bibfnamefont {E.~J.}\ \bibnamefont {Salumbides}},
  \bibinfo {author} {\bibfnamefont {U.}~\bibnamefont {Hollenstein}}, \bibinfo
  {author} {\bibfnamefont {J.~C.~J.}\ \bibnamefont {Koelemeij}}, \bibinfo
  {author} {\bibfnamefont {K.~S.~E.}\ \bibnamefont {Eikema}}, \bibinfo {author}
  {\bibfnamefont {W.}~\bibnamefont {Ubachs}}, \ and\ \bibinfo {author}
  {\bibfnamefont {F.}~\bibnamefont {Merkt}},\ }\href {\doibase
  10.1063/1.3120443} {\bibfield  {journal} {\bibinfo  {journal} {J. Chem.
  Phys.}\ }\textbf {\bibinfo {volume} {130}},\ \bibinfo {pages} {174306}
  (\bibinfo {year} {2009})}\BibitemShut {NoStop}%
\bibitem [{\citenamefont {Haase}\ \emph {et~al.}(2015)\citenamefont {Haase},
  \citenamefont {Beyer}, \citenamefont {Jungen},\ and\ \citenamefont
  {Merkt}}]{haase15a}%
  \BibitemOpen
  \bibfield  {author} {\bibinfo {author} {\bibfnamefont {C.}~\bibnamefont
  {Haase}}, \bibinfo {author} {\bibfnamefont {M.}~\bibnamefont {Beyer}},
  \bibinfo {author} {\bibfnamefont {C.}~\bibnamefont {Jungen}}, \ and\ \bibinfo
  {author} {\bibfnamefont {F.}~\bibnamefont {Merkt}},\ }\href {\doibase
  10.1063/1.4907531} {\bibfield  {journal} {\bibinfo  {journal} {J. Chem.
  Phys.}\ }\textbf {\bibinfo {volume} {142}},\ \bibinfo {pages} {064310}
  (\bibinfo {year} {2015})}\BibitemShut {NoStop}%
\bibitem [{\citenamefont {Wind}(1965)}]{wind65a}%
  \BibitemOpen
  \bibfield  {author} {\bibinfo {author} {\bibfnamefont {H.}~\bibnamefont
  {Wind}},\ }\href {\doibase 10.1063/1.1696302} {\bibfield  {journal} {\bibinfo
   {journal} {J. Chem. Phys.}\ }\textbf {\bibinfo {volume} {42}},\ \bibinfo
  {pages} {2371} (\bibinfo {year} {1965})}\BibitemShut {NoStop}%
\bibitem [{\citenamefont {Peek}(1965)}]{peek65a}%
  \BibitemOpen
  \bibfield  {author} {\bibinfo {author} {\bibfnamefont {J.~M.}\ \bibnamefont
  {Peek}},\ }\href {\doibase 10.1063/1.1697265} {\bibfield  {journal} {\bibinfo
   {journal} {J. Chem. Phys.}\ }\textbf {\bibinfo {volume} {43}},\ \bibinfo
  {pages} {3004} (\bibinfo {year} {1965})}\BibitemShut {NoStop}%
\bibitem [{\citenamefont {Kolos}(1969)}]{kolos69a}%
  \BibitemOpen
  \bibfield  {author} {\bibinfo {author} {\bibfnamefont {W.}~\bibnamefont
  {Kolos}},\ }\href {\doibase 10.1007/BF03156748} {\bibfield  {journal}
  {\bibinfo  {journal} {Acta Phys. Ac. Sc. Hung.}\ }\textbf {\bibinfo {volume}
  {27}},\ \bibinfo {pages} {241} (\bibinfo {year} {1969})}\BibitemShut
  {NoStop}%
\bibitem [{\citenamefont {Howells}\ and\ \citenamefont
  {Kennedy}(1990)}]{howells90a}%
  \BibitemOpen
  \bibfield  {author} {\bibinfo {author} {\bibfnamefont {M.~H.}\ \bibnamefont
  {Howells}}\ and\ \bibinfo {author} {\bibfnamefont {R.~A.}\ \bibnamefont
  {Kennedy}},\ }\href {\doibase 10.1039/ft9908603495} {\bibfield  {journal}
  {\bibinfo  {journal} {J. Chem. Soc., Faraday Trans.}\ }\textbf {\bibinfo
  {volume} {86}},\ \bibinfo {pages} {3495} (\bibinfo {year}
  {1990})}\BibitemShut {NoStop}%
\bibitem [{\citenamefont {Moss}\ and\ \citenamefont
  {Valenzano}(2003)}]{moss03a}%
  \BibitemOpen
  \bibfield  {author} {\bibinfo {author} {\bibfnamefont {R.~E.}\ \bibnamefont
  {Moss}}\ and\ \bibinfo {author} {\bibfnamefont {L.}~\bibnamefont
  {Valenzano}},\ }\href {\doibase 10.1080/00268970310001592683} {\bibfield
  {journal} {\bibinfo  {journal} {Mol. Phys.}\ }\textbf {\bibinfo {volume}
  {101}},\ \bibinfo {pages} {2635} (\bibinfo {year} {2003})}\BibitemShut
  {NoStop}%
\bibitem [{\citenamefont {M{\"u}ller-Dethlefs}\ and\ \citenamefont
  {Schlag}(1998)}]{muellerdethlefs98a}%
  \BibitemOpen
  \bibfield  {author} {\bibinfo {author} {\bibfnamefont {K.}~\bibnamefont
  {M{\"u}ller-Dethlefs}}\ and\ \bibinfo {author} {\bibfnamefont {E.~W.}\
  \bibnamefont {Schlag}},\ }\href {\doibase
  10.1002/(SICI)1521-3773(19980605)37:10<1346::AID-ANIE1346>3.0.CO;2-H}
  {\bibfield  {journal} {\bibinfo  {journal} {Angew. Chem. Int. Ed.}\ }\textbf
  {\bibinfo {volume} {37}},\ \bibinfo {pages} {1346} (\bibinfo {year}
  {1998})}\BibitemShut {NoStop}%
\bibitem [{\citenamefont {Hollenstein}\ \emph {et~al.}(2001)\citenamefont
  {Hollenstein}, \citenamefont {Seiler}, \citenamefont {Schmutz}, \citenamefont
  {Andrist},\ and\ \citenamefont {Merkt}}]{hollenstein01a}%
  \BibitemOpen
  \bibfield  {author} {\bibinfo {author} {\bibfnamefont {U.}~\bibnamefont
  {Hollenstein}}, \bibinfo {author} {\bibfnamefont {R.}~\bibnamefont {Seiler}},
  \bibinfo {author} {\bibfnamefont {H.}~\bibnamefont {Schmutz}}, \bibinfo
  {author} {\bibfnamefont {M.}~\bibnamefont {Andrist}}, \ and\ \bibinfo
  {author} {\bibfnamefont {F.}~\bibnamefont {Merkt}},\ }\href {\doibase
  10.1063/1.1396856} {\bibfield  {journal} {\bibinfo  {journal} {J. Chem.
  Phys.}\ }\textbf {\bibinfo {volume} {115}},\ \bibinfo {pages} {5461}
  (\bibinfo {year} {2001})}\BibitemShut {NoStop}%
\bibitem [{\citenamefont {Reinhold}\ \emph {et~al.}(1997)\citenamefont
  {Reinhold}, \citenamefont {Hogervorst},\ and\ \citenamefont
  {Ubachs}}]{reinhold97a}%
  \BibitemOpen
  \bibfield  {author} {\bibinfo {author} {\bibfnamefont {E.}~\bibnamefont
  {Reinhold}}, \bibinfo {author} {\bibfnamefont {W.}~\bibnamefont
  {Hogervorst}}, \ and\ \bibinfo {author} {\bibfnamefont {W.}~\bibnamefont
  {Ubachs}},\ }\href {\doibase 10.1103/PhysRevLett.78.2543} {\bibfield
  {journal} {\bibinfo  {journal} {Phys. Rev. Lett.}\ }\textbf {\bibinfo
  {volume} {78}},\ \bibinfo {pages} {2543} (\bibinfo {year}
  {1997})}\BibitemShut {NoStop}%
\bibitem [{\citenamefont {Reinhold}\ \emph {et~al.}(1999)\citenamefont
  {Reinhold}, \citenamefont {Hogervorst}, \citenamefont {Ubachs},\ and\
  \citenamefont {Wolniewicz}}]{reinhold99a}%
  \BibitemOpen
  \bibfield  {author} {\bibinfo {author} {\bibfnamefont {E.}~\bibnamefont
  {Reinhold}}, \bibinfo {author} {\bibfnamefont {W.}~\bibnamefont
  {Hogervorst}}, \bibinfo {author} {\bibfnamefont {W.}~\bibnamefont {Ubachs}},
  \ and\ \bibinfo {author} {\bibfnamefont {L.}~\bibnamefont {Wolniewicz}},\
  }\href {\doibase 10.1103/PhysRevA.60.1258} {\bibfield  {journal} {\bibinfo
  {journal} {Phys. Rev. A}\ }\textbf {\bibinfo {volume} {60}},\ \bibinfo
  {pages} {1258} (\bibinfo {year} {1999})}\BibitemShut {NoStop}%
\bibitem [{\citenamefont {Wolniewicz}(1998)}]{wolniewicz98b}%
  \BibitemOpen
  \bibfield  {author} {\bibinfo {author} {\bibfnamefont {L.}~\bibnamefont
  {Wolniewicz}},\ }\href {\doibase 10.1063/1.475521} {\bibfield  {journal}
  {\bibinfo  {journal} {J. Chem. Phys.}\ }\textbf {\bibinfo {volume} {108}},\
  \bibinfo {pages} {1499} (\bibinfo {year} {1998})}\BibitemShut {NoStop}%
\bibitem [{\citenamefont {Merkt}\ \emph {et~al.}(1998)\citenamefont {Merkt},
  \citenamefont {Osterwalder}, \citenamefont {Seiler}, \citenamefont
  {Signorell}, \citenamefont {Palm}, \citenamefont {Schmutz},\ and\
  \citenamefont {Gunzinger}}]{merkt98a}%
  \BibitemOpen
  \bibfield  {author} {\bibinfo {author} {\bibfnamefont {F.}~\bibnamefont
  {Merkt}}, \bibinfo {author} {\bibfnamefont {A.}~\bibnamefont {Osterwalder}},
  \bibinfo {author} {\bibfnamefont {R.}~\bibnamefont {Seiler}}, \bibinfo
  {author} {\bibfnamefont {R.}~\bibnamefont {Signorell}}, \bibinfo {author}
  {\bibfnamefont {H.}~\bibnamefont {Palm}}, \bibinfo {author} {\bibfnamefont
  {H.}~\bibnamefont {Schmutz}}, \ and\ \bibinfo {author} {\bibfnamefont
  {R.}~\bibnamefont {Gunzinger}},\ }\href {\doibase 10.1088/0953-4075/31/8/020}
  {\bibfield  {journal} {\bibinfo  {journal} {J. Phys. B}\ }\textbf {\bibinfo
  {volume} {31}},\ \bibinfo {pages} {1705} (\bibinfo {year}
  {1998})}\BibitemShut {NoStop}%
\bibitem [{\citenamefont {Zhu}\ and\ \citenamefont {Johnson}(1991)}]{zhu91a}%
  \BibitemOpen
  \bibfield  {author} {\bibinfo {author} {\bibfnamefont {L.}~\bibnamefont
  {Zhu}}\ and\ \bibinfo {author} {\bibfnamefont {P.}~\bibnamefont {Johnson}},\
  }\href {\doibase 10.1063/1.460460} {\bibfield  {journal} {\bibinfo  {journal}
  {J. Chem. Phys.}\ }\textbf {\bibinfo {volume} {94}},\ \bibinfo {pages} {5769}
  (\bibinfo {year} {1991})}\BibitemShut {NoStop}%
\bibitem [{\citenamefont {Mohr}\ \emph {et~al.}()\citenamefont {Mohr},
  \citenamefont {Taylor},\ and\ \citenamefont {Newell}}]{mohr14a}%
  \BibitemOpen
  \bibfield  {author} {\bibinfo {author} {\bibfnamefont {P.~J.}\ \bibnamefont
  {Mohr}}, \bibinfo {author} {\bibfnamefont {B.~N.}\ \bibnamefont {Taylor}}, \
  and\ \bibinfo {author} {\bibfnamefont {D.~B.}\ \bibnamefont {Newell}},\
  }\href@noop {} {\emph {\bibinfo {title} {The 2014 CODATA Recommended Values
  of the Fundamental Physical Constants (Web Version 7.0). 
  http://physics.nist.gov/constants (2015)}}}.\ \BibitemShut {Stop}%
\bibitem [{\citenamefont {Xie}\ and\ \citenamefont {Zare}(1990)}]{xie90a}%
  \BibitemOpen
  \bibfield  {author} {\bibinfo {author} {\bibfnamefont {J.}~\bibnamefont
  {Xie}}\ and\ \bibinfo {author} {\bibfnamefont {R.~N.}\ \bibnamefont {Zare}},\
  }\href {\doibase 10.1063/1.458837} {\bibfield  {journal} {\bibinfo  {journal}
  {J. Chem. Phys.}\ }\textbf {\bibinfo {volume} {93}},\ \bibinfo {pages} {3033}
  (\bibinfo {year} {1990})}\BibitemShut {NoStop}%
\bibitem [{\citenamefont {Signorell}\ and\ \citenamefont
  {Merkt}(1997)}]{signorell97a}%
  \BibitemOpen
  \bibfield  {author} {\bibinfo {author} {\bibfnamefont {R.}~\bibnamefont
  {Signorell}}\ and\ \bibinfo {author} {\bibfnamefont {F.}~\bibnamefont
  {Merkt}},\ }\href {\doibase 10.1080/002689797169745} {\bibfield  {journal}
  {\bibinfo  {journal} {Mol. Phys.}\ }\textbf {\bibinfo {volume} {92}},\
  \bibinfo {pages} {793} (\bibinfo {year} {1997})}\BibitemShut {NoStop}%
\bibitem [{\citenamefont {Johnson}(1977)}]{johnson77a}%
  \BibitemOpen
  \bibfield  {author} {\bibinfo {author} {\bibfnamefont {B.~R.}\ \bibnamefont
  {Johnson}},\ }\href {\doibase 10.1063/1.435384} {\bibfield  {journal}
  {\bibinfo  {journal} {J. Chem. Phys.}\ }\textbf {\bibinfo {volume} {67}},\
  \bibinfo {pages} {4086} (\bibinfo {year} {1977})}\BibitemShut {NoStop}%
\bibitem [{\citenamefont {Wolniewicz}(1966)}]{wolniewicz66a}%
  \BibitemOpen
  \bibfield  {author} {\bibinfo {author} {\bibfnamefont {L.}~\bibnamefont
  {Wolniewicz}},\ }\href {\doibase 10.1063/1.1727599} {\bibfield  {journal}
  {\bibinfo  {journal} {J. Chem. Phys.}\ }\textbf {\bibinfo {volume} {45}},\
  \bibinfo {pages} {515} (\bibinfo {year} {1966})}\BibitemShut {NoStop}%
\bibitem [{\citenamefont {Teller}\ and\ \citenamefont
  {Sahlin}(1970)}]{teller70a}%
  \BibitemOpen
  \bibfield  {author} {\bibinfo {author} {\bibfnamefont {E.}~\bibnamefont
  {Teller}}\ and\ \bibinfo {author} {\bibfnamefont {H.~L.}\ \bibnamefont
  {Sahlin}},\ }in\ \href@noop {} {\emph {\bibinfo {booktitle} {Physical
  Chemistry: An Advanced Treatise}}},\ Vol.~\bibinfo {volume} {V}\ (\bibinfo
  {publisher} {Academic Press},\ \bibinfo {address} {New York},\ \bibinfo
  {year} {1970})\ Chap.~\bibinfo {chapter} {1}\BibitemShut {NoStop}%
\bibitem [{\citenamefont {Smith}(1971)}]{smith71a}%
  \BibitemOpen
  \bibfield  {author} {\bibinfo {author} {\bibfnamefont {K.}~\bibnamefont
  {Smith}},\ }\enquote {\bibinfo {title} {The calculation of atomic collision
  processes},}\ \ (\bibinfo  {publisher} {Wiley},\ \bibinfo {address} {New
  York},\ \bibinfo {year} {1971})\ Chap.~\bibinfo {chapter} {1}\BibitemShut
  {NoStop}%
\end{thebibliography}

\end{document}